\documentclass[prc,twocolumn,nofootinbib,superscriptaddress]{revtex4-1}

\usepackage{graphicx,amsmath,amssymb,bm,tabularx}
\usepackage{multirow,dcolumn}

\newcommand {\Fig} [1] {Fig.~\ref{#1}}
\newcommand {\Eq} [1] {Eq.~(\ref{#1})}

\newcolumntype{.}{D{.}{.}{2.3}}

\begin{document}

\title{Double-folding potentials from chiral effective field theory}

\author{V.\ Durant}
\email[Email:~]{durant@theorie.ikp.physik.tu-darmstadt.de}
\affiliation{Institut f\"ur Kernphysik, Technische Universit\"at Darmstadt, 64289 Darmstadt, Germany}
\affiliation{ExtreMe Matter Institute EMMI, GSI Helmholtzzentrum f\"ur Schwerionenforschung GmbH, 64291 Darmstadt, Germany}

\author{P.\ Capel}
\email[Email:~]{pierre.capel@ulb.ac.be}
\affiliation{Institut f\"ur Kernphysik, Technische Universit\"at Darmstadt, 64289 Darmstadt, Germany}
\affiliation{ExtreMe Matter Institute EMMI, GSI Helmholtzzentrum f\"ur Schwerionenforschung GmbH, 64291 Darmstadt, Germany}
\affiliation{Physique Nucl\'eaire et Physique Quantique (CP 229),Universit\'e libre de Bruxelles (ULB), B-1050 Brussels, Belgium}

\author{L.\ Huth}
\email[Email:~]{lukashuth@theorie.ikp.physik.tu-darmstadt.de}
\affiliation{Institut f\"ur Kernphysik, Technische Universit\"at Darmstadt, 64289 Darmstadt, Germany}
\affiliation{ExtreMe Matter Institute EMMI, GSI Helmholtzzentrum f\"ur Schwerionenforschung GmbH, 64291 Darmstadt, Germany}

\author{A. B. Balantekin}
\email[Email:~]{baha@physics.wisc.edu}
\affiliation{Department of Physics, University of Wisconsin, Madison, WI 53706, USA}

\author{A.\ Schwenk}
\email[Email:~]{schwenk@physik.tu-darmstadt.de}
\affiliation{Institut f\"ur Kernphysik, Technische Universit\"at Darmstadt, 64289 Darmstadt, Germany}
\affiliation{ExtreMe Matter Institute EMMI, GSI Helmholtzzentrum f\"ur Schwerionenforschung GmbH, 64291 Darmstadt, Germany}
\affiliation{Max-Planck-Institut f\"ur Kernphysik, Saupfercheckweg 1, 69117 Heidelberg, Germany}

\begin{abstract}
The determination of nucleus-nucleus potentials is important not only
to describe the properties of the colliding system, but also to
extract nuclear-structure information and for modelling nuclear
reactions for astrophysics. We present the first determination of
double-folding potentials based on chiral effective field theory at
leading, next-to-leading, and next-to-next-to-leading order. To this
end, we construct new soft local chiral effective field theory
interactions. We benchmark this approach in the $^{16}$O--$^{16}$O
system, and present results for cross sections computed for elastic
scattering up to 700 MeV in energy, as well as for the astrophysical
$S$-factor of the fusion reaction.
\end{abstract}

\maketitle

\section{Introduction}

Determining the interaction between two nuclei is a long-standing and
challenging problem~\cite{Bran97heavyion}. It constitutes an important
input in the modelling of nuclear reactions, which provide key
information about the structure of nuclei and are relevant for
processes that take place in stars.  The interaction between two
nuclei has been modelled by phenomenological potentials, e.g., of
Woods-Saxon form, whose parameters are adjusted to reproduce
elastic-scattering data. Numerical potentials have also been obtained
from inversion of scattering data~\cite{Leeb85PotInvdata}.  Albeit
precise when experimental data exist, these potentials lack predictive
behavior and do not have controlled uncertainties.  Alternatively, it
has been suggested to construct nucleus-nucleus potentials from the
densities of the colliding nuclei and a given nucleon-nucleon ($NN$)
interaction using a double-folding procedure~\cite{Satc79Folding}. It
is known that this framework provides more realistic potentials for
the nucleon-nucleus interactions than for the nucleus-nucleus
case~\cite{Maha91eqpot}. Nevertheless, it constitutes a first-order
approximation to optical potentials derived from Feshbach's reaction
theory~\cite{Bran97heavyion}.  Interesting results have been obtained
in such a way, e.g., by considering zero-range contact $NN$
interactions~\cite{Cham02SPdens,Pere09ImDFPot} or using a $G$-matrix
approach, see, e.g., Refs.~\cite{Furu12OpPot,Mino15ChEFTCC} for recent work.

During the last decade, there have been great advances in nuclear
structure and nuclear reactions based on effective field theories
(EFT). For example, chiral EFT has become the standard method for
developing systematic nuclear forces rooted in the symmetries of
quantum chromodynamics (see, e.g., Refs.~\cite{Epel09RMP,Mach11PR,Hamm13RMP}
for recent reviews).  Efforts have been made to derive nucleon-nucleus
optical potentials using chiral EFT interactions from many-body
perturbation theory~\cite{Holt13OpPotChEFT,Vor15OpPotChEFT} and
self-consistent Green's function calculations~\cite{Rot17,Idi16}. In
this work, we focus on the derivation of nucleus-nucleus potentials
from chiral $NN$ interactions and nucleonic densities. In particular,
we use local chiral EFT
interactions~\cite{Geze13QMCchi,Geze14long,Lynn14QMCln,Tews16QMCPNM,Lynn16QMC3N,Lynn17QMClight,Huth17Fierz},
because this simplifies the double-folding calculation.

In this first study, we explore and test this idea for
$^{16}$O--$^{16}$O reactions, comparing our calculations to
elastic-scattering~\cite{Stil89expel,Bohl93expel,Sugi93expel,Kond96expel,Bart96expel,Nuof98expel,Nico99expel,Khoa0016OEl}
and fusion
data~\cite{Tser78expfus,Hulk80expfus,Wu84expfus,Kuro87expfus,Duar15expfus}. For
this system, phenomenological Woods-Saxon
potentials~\cite{Khoa0016OEl} and potentials obtained through
inversion techniques~\cite{Alle93OpPotInv16O,Benn96InvSctdat} also
exist.  We show that the double-folding potential and the reaction
observables exhibit an order-by-order behavior expected in EFT and
observe that, for soft potentials, our calculations have only a weak
dependence on the regularization scale.  The comparison of our results
with experiment leads us to suggest various directions for
improvements for constructing nucleus-nucleus potentials from chiral
EFT interactions with the double-folding method.

This paper is organized as follows. We start with a brief review of
the formalism for the double-folding potential in the following
section.  In Sec.~\ref{sec:local_cheft}, we discuss local chiral EFT
interactions and the construction of new soft local chiral $NN$
potentials. We then determine the double-folding potentials at
different chiral orders and apply these to $^{16}$O--$^{16}$O elastic
scattering in Sec.~\ref{sec:elastic} and to the $S$-factor for
$^{16}$O+$^{16}$O fusion in Sec.~\ref{sec:S_fac}. Finally, we
summarize and give an outlook in Sec. \ref{sec:sum}.

\section{Double-folding potential: Formalism}
\label{sec:formalism}

We consider the potential between nucleus 1 (with atomic and mass
numbers $Z_1$ and $A_1$) and nucleus 2 (with $Z_2$ and $A_2$). In the
double-folding formalism, the nuclear part of the nucleus-nucleus
potential $V_\text{F}=V_\text{D}+V_\text{Ex}$ can be constructed from
a given $NN$ interaction $v$ by double folding over the densities in
the direct (D) channel and the density matrices in the exchange (Ex)
channel. The review of the formalism for the double-folding potential
in this section follows Ref.~\cite{Furu12OpPot}. We will include only
$NN$ interactions here and leave the investigation of many-body
contributions to future work.

\begin{figure}[t]
\begin{center}
\includegraphics[width=0.75\columnwidth]{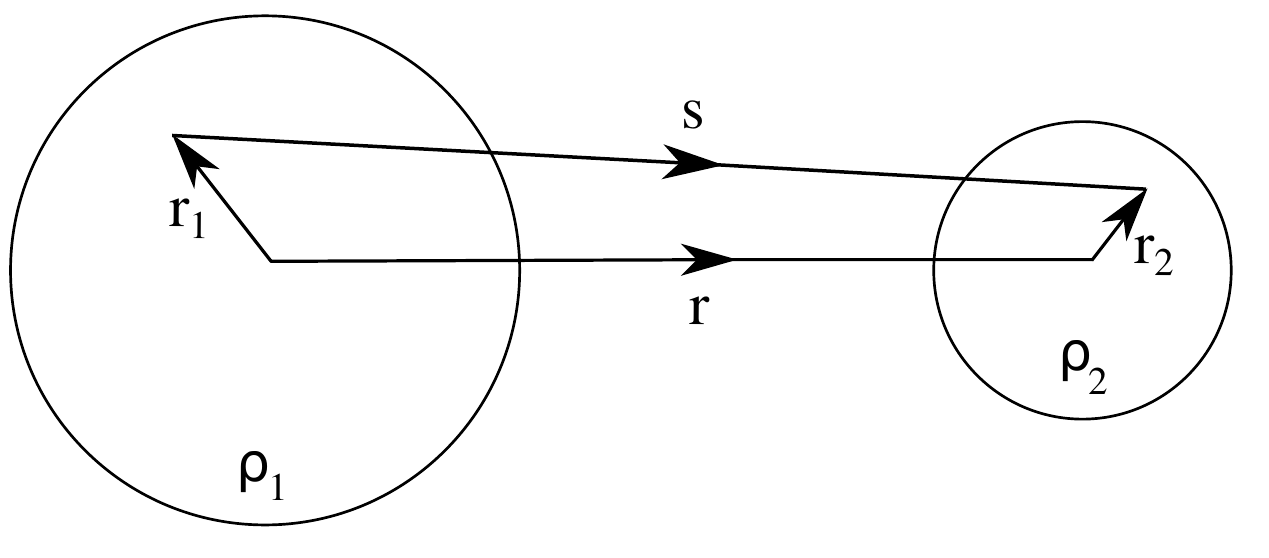}
\end{center}
\caption{\label{fig:coordinates}
Coordinates of the nuclei involved in the double-folding calculation.}
\end{figure}

In the direct channel, the double-folding potential is calculated by
integrating the $NN$ interaction over the neutron ($n$) and proton
($p$) density distributions $\rho_1^{n,p}$ and $\rho_2^{n,p}$ of the
colliding nuclei,
\begin{equation}
V_\text{D}({\bf r}) = \sum_{i,j = n,p} \int \int \rho^i_1({\bf r}_1) \, v^{ij}_\text{D}({\bf s}) \, \rho^j_2({\bf r}_2)
\, d^3{\bf r}_1 d^3{\bf r}_2 \,,
\label{eq:direct}
\end{equation}
where ${\bf r}$ is the relative coordinate between the center of mass
of the nuclei, ${\bf r}_1$ and ${\bf r}_2$ are the coordinates from
the center of mass of each nucleus, ${\bf s}={\bf r}-{\bf r}_1+{\bf
r}_2$ (the geometry is shown in \Fig{fig:coordinates}), and the sum
$i,j$ is over neutrons and protons with their respective densities.

To account for the antisymmetrization between nucleons, the
double-folding potential receives contributions also from the exchange channel,
\begin{multline}
V_\text{Ex}({\bf r},E_\text{cm}) = \sum_{i,j = n,p} \int \int \rho^i_1({\bf r}_1,{\bf r}_1+{\bf s})
\, v^{ij}_\text{Ex}({\bf s}) \\
\times \rho^j_2({\bf r}_2,{\bf r}_2-{\bf s}) \exp \left[\frac{i{\bf k}({\bf r})\cdot{\bf s}}{\mu/m_N}\right] \,
d^3{\bf r}_1 d^3{\bf r}_2 \,, \label{eq:exchange}
\end{multline}
where $\mu=m_NA_1A_2/(A_1+A_2)$ is the reduced mass of the colliding
nuclei (with $m_N$ the nucleon mass) and the integral is over the
density matrices $\rho^i({\bf r},{\bf r} \pm {\bf s})$ of the
nuclei. In the exchange channel, there is an additional phase that
renders the double-folding potential dependent on the energy
$E_\text{cm}$ in the center-of-mass system. The momentum for the
nucleus-nucleus relative motion ${\bf k}$ is related to $E_\text{cm}$,
the nuclear part of the double-folding potential, and the
double-folding Coulomb potential $V_\text{Coul}$ through
\begin{equation}
k^2({\bf r})=2\mu \, \Bigl[ E_\text{cm} - V_\text{F}({\bf r},E_\text{cm}) - V_\text{Coul}({\bf r}) 
\Bigr] \,. \label{eq:k}
\end{equation}
As a result, $V_\text{Ex}$ has to be determined self-consistently.
Note that at our level of calculation the double-folding potential,
$V_\text{F}=V_\text{D}+V_\text{Ex}$, is real.  The density matrices entering in
\Eq{eq:exchange} are approximated using the density matrix
expansion~\cite{Nege72DME1} restricted to its leading term,
\begin{equation}
\rho^i({\bf r},{\bf r} \pm {\bf s}) = \frac{3}{s\, k^i_\text{F}({\bf R})} \, j_1\bigr(s\, k^i_\text{F}({\bf R})
\bigl) \, \rho^i({\bf R}) \,,
\end{equation}
where ${\bf R} = {\bf r} \pm {\bf s}/2$, $j_1$ is a spherical Bessel function
of the first kind, and we take the effective local Fermi momentum, which is an
arbitrary scale in the density-matrix expansion, as in Ref.~\cite{Furu12OpPot}:
\begin{equation}
k^i_\text{F} = \biggl[ (3 \pi^2 \rho^i)^{2/3} + \frac{5 (\nabla \rho^i)^2}{12 (\rho^i)^2}
+ \frac{5 \nabla^2 \rho^i}{36 \rho^i} \biggr]^{1/2} \,.
\end{equation}
In the case of spherical nuclei, the densities and the effective local Fermi
momenta depend only on the distance from the center of mass of the nucleus
($r_i$ or $R$).

For doubly closed-shell nuclei, the $NN$ interaction entering the
double-folding potential in the direct and exchange channels,
$v_\text{D}$ and $v_\text{Ex}$, respectively, at this level receive
contributions only from the central parts of nuclear forces. Then also
the $NN$ interaction and the double-folding potentials depend only on
the relative distance ($s$ or $r$). Writing the $NN$ interaction in
terms of their two-body spin-isospin components, $v^{ST}$, and
distinguishing between proton-proton ($pp$), proton-neutron ($pn,np$),
and neutron-neutron ($nn$) interactions, $v_\text{D}$ and
$v_\text{Ex}$ read
\begin{align}
v^{pp,nn}_\text{D,Ex}(s) &= \frac{1}{4} \Bigl[v^{01}(s) \pm 3v^{11}(s)\Bigr] \,, \\
v^{pn,np}_{\rm D,Ex}(s) &= \frac{1}{8} \Bigl[\pm v^{00}(s) + v^{01}(s) + 3v^{10}(s) \pm 3v^{11}(s)\Bigr] \,,
\end{align}
where the upper (lower) signs refer to the direct (exchange) term and
we have neglected the small isospin-symmetry-breaking corrections to
$v$.

The densities of the colliding nuclei are an important input for the
calculation of the double-folding potential. In this first study based
on chiral EFT interactions, we adopt the two-parameter Fermi
distributions provided by the S\~ao Paulo group~\cite{Cham02SPdens}
for the proton and neutron densities, whose parameters were fitted to
Dirac-Hartree-Bogoliubov calculations
\begin{equation}
\rho^{p,n}(r)=\frac{\rho_0}{1+\exp\left(\frac{r-R_{p,n}}{a_{p,n}}\right)} \,,
\end{equation}
where $\rho_0=0.091$~fm$^{-3}$ and the radii $R_{p,n}$ and
diffusenesses $a_{p,n}$ depend on the proton and neutron numbers of
the nucleus. Expressed in fm, they are given by
\begin{align}
R_p &= 1.81 \, Z^{1/3} -1.12 \,, \quad a_p=0.47-0.00083 \, Z \,, \\
R_n &= 1.49 \, N^{1/3} -0.79 \,, \quad a_n=0.47+0.00046 \, N \,.
\end{align}

\section{Local chiral EFT interactions}
\label{sec:local_cheft}

\subsection{Nucleon-nucleon potentials} 

Chiral EFT provides a systematic expansion for nuclear forces using
nucleons and pions as degrees of freedom, which is connected to the
underlying theory of quantum
chromodynamics~\cite{Epel09RMP,Mach11PR}. The different contributions
to $NN$ and many-nucleon interactions are ordered according to a power
counting scheme in powers of $(Q/\Lambda_b)^\nu$, where $Q$ is a
typical momentum or the pion mass and $\Lambda_b$ the breakdown scale
of the theory of the order of 500 MeV. This leads to a hierarchy of
two- and many-nucleon interactions, with $NN$ interactions starting at
leading order (LO, $\nu = 0$) followed by a contribution at
next-to-leading order (NLO, $\nu=2$), whereas three-nucleon
interactions enter at next-to-next-to leading order (N$^2$LO, $\nu =
3$).

Because they facilitate the calculation of double-folding potentials,
we use local chiral $NN$ interactions, developed initially in
Refs.~\cite{Geze13QMCchi,Geze14long}, but construct new soft $NN$
interactions up to N$^2$LO.  As in these references, the long- and
short-range parts of the interaction are regularized by
\begin{equation}
f_\text{long}(r) = 1 - e^{-(r/R_0)^4} \hspace{1.5mm} \text{and} \hspace{2mm} f_\text{short}(r) = \frac{e^{-(r/R_0)^4}}{\pi \Gamma(3/4) R_0^3} \,,
\end{equation}
where $R_0$ is the coordinate-space cutoff in the $NN$ potentials
used. The long-range regulator is designed to remove the singularity
at $r=0$ in the pion exchanges, while it preserves its properties at
large distances. The short-range regulator smears out the $NN$ contact
interactions.  A second cutoff $\tilde{\Lambda}$ is used in the
spectral-function regularization of the two-pion exchange, which
enters first at NLO.  In Refs.~\cite{Geze13QMCchi,Geze14long} it was
shown that the calculations are practically insensitive to
$\tilde{\Lambda}$ for local interactions; in the present work, we
consider $\tilde{\Lambda}=1000$~MeV.

\begin{table}[t]
\caption{Low-energy constants (LECs) for the soft cutoffs $R_0=1.4$~fm
  and 1.6~fm at LO, NLO and N$^2$LO. In the last row, the deuteron
  binding energy $E_d$, not used to constrain the LECs, is given in
  MeV; its experimental value is $E_d = 2.224$~MeV.  The LO LECs
  $C_{S}$ and $C_{T}$ are given in fm$^2$, the others are in fm$^4$.}
\label{tab:LECs}
{\renewcommand{\arraystretch}{1.3}
\begin{ruledtabular}
\begin{tabular}{c|...|...}
$R_0$ [fm] & & \multicolumn{1}{c}{1.4~fm} & & & \multicolumn{1}{c}{1.6~fm} & \\
& \multicolumn{1}{c}{LO} & \multicolumn{1}{c}{NLO} & \multicolumn{1}{c|}{N$^2$LO} & \multicolumn{1}{c}{LO} & \multicolumn{1}{c}{NLO} & \multicolumn{1}{c}{N$^2$LO} \\
\hline
$C_S$ & -2.675 & -0.480 & 1.331 & -3.590 & -0.988 & 0.538 \\
$C_T$ & -0.021 & 0.723 & 0.363 & -0.188 & 0.660 & 0.495 \\
$C_1$ & & 0.124 & -0.104 & & -0.098 & -0.206 \\
$C_2$ & & 0.302 & 0.188 & & 0.393 & 0.368 \\
$C_3$ & & -0.224 & -0.217 & & -0.311 & -0.278 \\
$C_4$ & & 0.192 & 0.166 & & 0.312 & 0.267 \\
$C_5$ & & -2.268 & -2.083 & & -2.416 & -2.282 \\
$C_6$ & & 0.471 & 0.355 & & 0.603 & 0.536 \\
$C_7$ & & -0.578 & -0.529 & & -0.798 & -0.790 \\
\hline
$E_d$ & 1.886 & 2.151 & 2.193 & 2.043 & 2.147 & 2.178
\end{tabular}
\end{ruledtabular}}
\end{table}

\begin{figure}[t]
\includegraphics[width=\columnwidth]{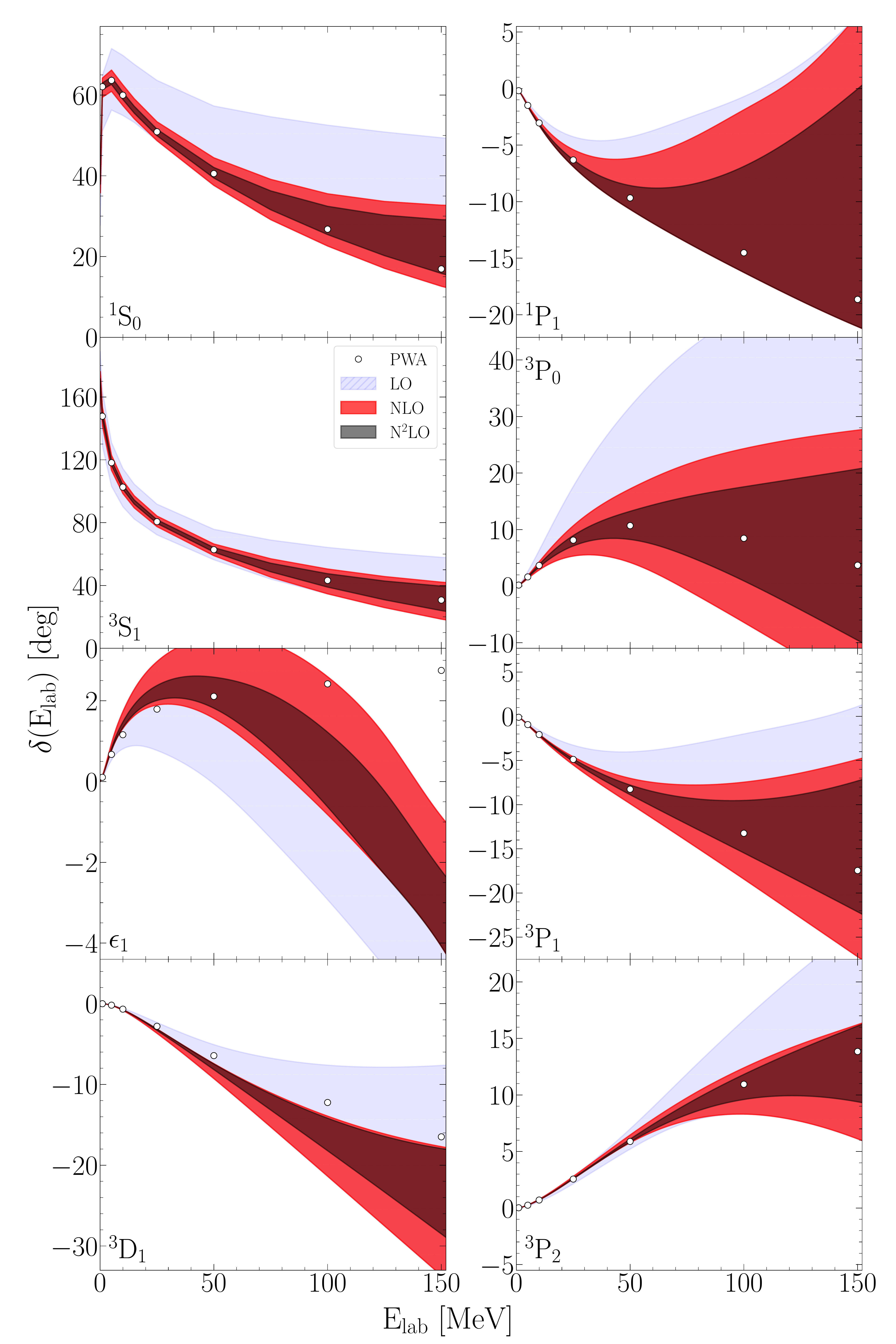}
\caption{Phase shifts for $R_0=1.4$~fm in different partial waves as a
  function of laboratory energy.  Results are shown for the LO (blue),
  NLO (red), and N$^2$LO (grey) interactions compared to the Nijmegen
  partial wave analysis (PWA)~\cite{Stok93PWA}. The bands at each
  order give the theoretical uncertainty as discussed in the text.}
\label{fig:Phase_shifts}
\end{figure}

As it turns out, the available local interactions from
Refs.~\cite{Geze13QMCchi,Geze14long} with $R_0=1.0$~fm and 1.1~fm are
too hard (see also Ref.~\cite{Hopp17WeinEV}) and, thus, not suitable
for calculations of a nucleus-nucleus potential at the simple
``Hartree-Fock'' level\footnote{The double-folding potential is
  calculated at the Hartree-Fock level, but using phenomenological
  densities, which would otherwise be deficient, when taking them from
  a Hartree-Fock calculation based on $NN$ interactions.} considered
here, because the resulting double-folding potentials are
repulsive. Additional $NN$ attraction coming from beyond Hartree-Fock
many-body contributions would solve this behavior of the resulting
double-folding potentials. To perform calculations at the Hartree-Fock
level, we can only use the existing interaction with $R_0=1.2$~fm. In
order to estimate the impact of the regulator, we construct softer
interactions with cutoffs $R_0 =1.4$~fm and 1.6~fm. We determine the
low-energy constants (LECs) by fitting to the $np$ phase shifts from
the Nijmegen partial wave analysis (PWA)~\cite{Stok93PWA}.  To this
end, we minimize the following $\chi^2$
\begin{equation}
\chi^2 = \sum_i \frac{(\delta^\mathrm{PWA}_i - \delta^\mathrm{theo}_i)^2}{ \Delta \delta_i^2} \,,
\end{equation}
computed from the squared difference between the PWA phase shifts and
the calculated ones.  The uncertainty $\Delta \delta_i^2 $ is obtained
from the PWA, a model uncertainty, and a numerical error:
\begin{equation}
\Delta \delta_i^2  = (\Delta\delta_i^\mathrm{PWA})^2 + (\Delta\delta_i^\mathrm{mod})^2 +(\Delta\delta_i^\mathrm{num})^2 \,.
\end{equation}
For the model uncertainty we use a relative uncertainty multiplied
with a constant value~\cite{Carl15sim,Huth17Fierz},
\begin{align}
\Delta \delta_i^{\text{model, LO}} &= \left(\frac{Q}{\Lambda_b}\right)^2 C \,, 
\label{eq:loerr} \\
\Delta \delta_i^{\text{model, }\nu} &= \left(\frac{Q}{\Lambda_b}\right)^{\nu+1} C \,,
\label{eq:n2loerr}
\end{align}
where $Q=\text{max}(m_\pi,\,p = \sqrt{E^\mathrm{lab}_i m_N/2})$ and
$C= 1^\circ$.  For both cutoffs ($R_0 = 1.4$~fm and 1.6~fm), we take
$\Lambda_b = 400$~MeV, which also roughly corresponds to a
coordinate-space cutoff $R_0=1.4$~fm to get a more conservative
uncertainty estimate.

Our interactions are fit up to laboratory energies of 50~MeV at LO and
up to 150~MeV at NLO and N$^2$LO. In particular, we consider the
energies 1, 5, 10, 25, 50, 100, and 150 MeV. The LO interaction is fit
to the two $S$-wave channels, while the NLO and N$^2$LO interactions
are also constrained by the four $P$-waves and the $^3S_1$--$^3D_1$
mixing angle $\varepsilon_1$.  The LECs and the deuteron binding
energy obtained for each interaction are given in
Table~\ref{tab:LECs}.  All other inputs and conventions for these
softer local chiral $NN$ potentials are as in
Refs.~\cite{Geze13QMCchi,Geze14long}.  The phase shifts for
$R_0=1.4$~fm are shown in Fig.~\ref{fig:Phase_shifts}; we find similar
results with $R_0=1.6$~fm.  The phase shift reproduction here is
comparable to the interactions from
Refs.~\cite{Geze13QMCchi,Geze14long}.

\subsection{Double-folding potential}

To apply the double-folding method using local chiral $NN$
interactions, we consider the $^{16}$O--$^{16}$O system, where there
are ample sets of data to which we can compare our
calculations. Elastic scattering has been accurately measured at
various
energies~\cite{Stil89expel,Bohl93expel,Sugi93expel,Kond96expel,Bart96expel,Nuof98expel,Khoa0016OEl,Nico99expel}
and these data sets have been precisely analyzed with phenomenological
optical potentials~\cite{Khoa0016OEl,Kond9016ODeepPot} or using
inversion techniques~\cite{Alle93OpPotInv16O,Benn96InvSctdat}. This
enables us to compare our results with state-of-the-art
phenomenological calculations. At lower energy, the fusion of two
$^{16}$O
nuclei~\cite{Tser78expfus,Hulk80expfus,Wu84expfus,Kuro87expfus,Duar15expfus}
is another observable with which we can test our double-folding
potential. In this section, we present results for the double-folding
potential computed at different energies and we illustrate its
order-by-order behavior and the sensitivity to the cutoff scale.

\begin{figure}[t]
\includegraphics[width=\columnwidth]{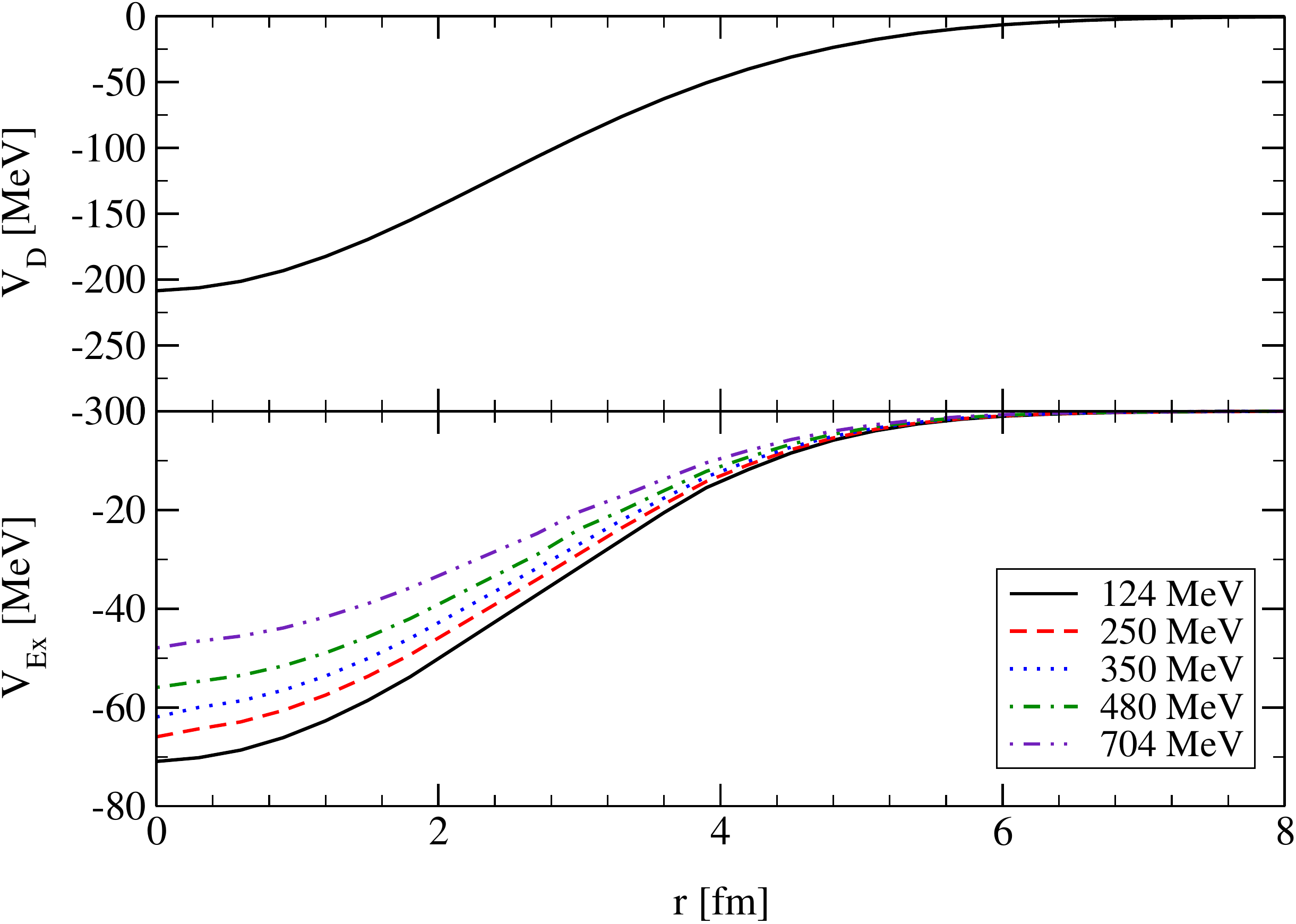}
\caption{Direct (upper panel) and exchange (lower panel) contributions
  to the double-folding potential for the $^{16}$O--$^{16}$O system
  based on the local chiral EFT interaction at N$^2$LO with
  $R_0=1.4$~fm. The direct contribution is energy independent and we
  show results for different laboratory energies, $E_\text{lab}$, in
  the exchange channel.}
\label{fig:D+Ex}
\end{figure}

\begin{figure}[t]
\includegraphics[width=\columnwidth]{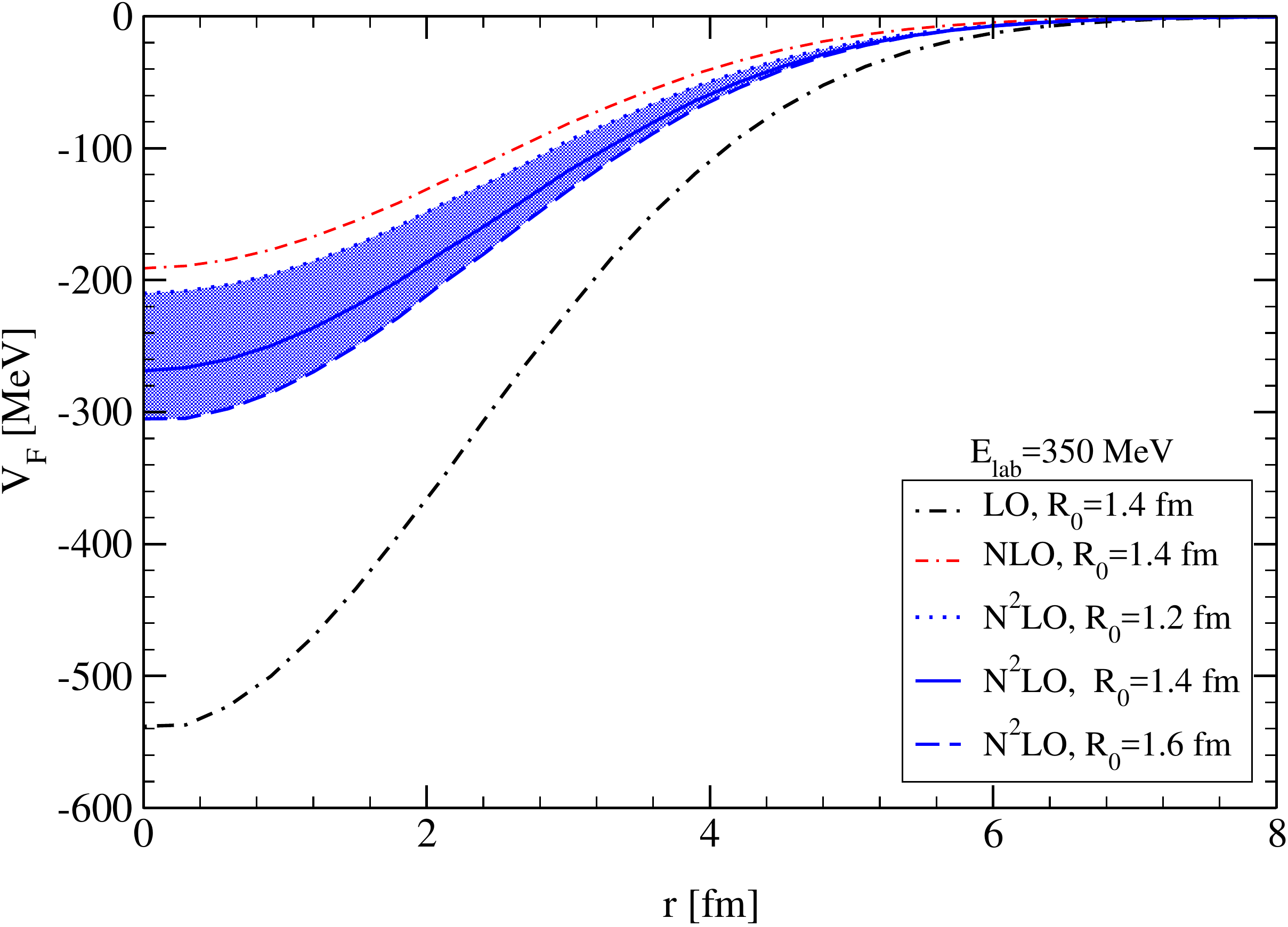}
\caption{Double-folding potential for the $^{16}$O--$^{16}$O system at
  $E_{\text{lab}}=350$~MeV.  The results obtained at LO, NLO, and
  N$^2$LO (for $R_0=1.4$~fm) illustrate the order-by-order behavior,
  while the shaded area at N$^2$LO shows the sensitivity to the cutoff for $R_0=1.2$, 1.4, and
  1.6~fm.}
\label{fig:VF_350}
\end{figure}

Figure~\ref{fig:D+Ex} shows the direct (upper panel) and exchange
(lower panel) contributions to the double-folding potential based on
the local chiral N$^2$LO potential with $R_0 = 1.4$~fm. Since the $NN$
interaction is energy independent, the direct contribution of the
double-folding potential is also energy independent [see
\Eq{eq:direct}]. The exchange contribution given by
\Eq{eq:exchange}, however, includes an energy dependence through the
relative momentum ${\bf k}$ in the exponential factor [see
\Eq{eq:k}]. The shape of this exchange contribution does not vary
significantly with energy, but its attractive strength decreases with
increasing energy, which can be understood by the increasing variation
of the exponential factor.

The final double-folding potential computed at different orders and
with different cutoffs is displayed for $E_\text{lab}=350$~MeV in
\Fig{fig:VF_350}. The order-by-order behavior is similar to what is
observed in \Fig{fig:Phase_shifts}.  As explained before, lower
cutoffs ($R_0 < 1.2$~fm) provide harder $NN$ interactions, which lead
to repulsive double-folding potentials at LO and NLO.  These
interactions require the additional attraction expected to come from
many-body contributions beyond the simple Hartree-Fock level
considered here.  At N$^2$LO, the calculations have been performed
with three different $NN$ cutoffs: $R_0=1.2$~fm (dotted line), 1.4~fm
(solid line), and 1.6~fm (dashed line); the lowest cutoff providing
the less attractive potential. It is interesting to notice that the
sensitivity to the $NN$ cutoff $R_0$ decreases at larger distance,
where all three N$^2$LO potentials present nearly identical
asymptotics. The range of the regularization cutoff, $R_0$,
highlighted by the shaded band in \Fig{fig:VF_350}, will allow us to
gauge the level of details needed in $NN$ interactions to reproduce
the physical observables in nucleus-nucleus reactions.

\section{Elastic scattering}
\label{sec:elastic}

The elastic scattering of medium to heavy nuclei can be described
within the optical model.  In that model, the nuclear part of the
interaction between the colliding nuclei is described by a complex
potential. Roughly speaking, the real part corresponds to the
attractive interaction between the nuclei, whereas the imaginary part
simulates the absorption of the incoming channel to other open
channels, such as inelastic scattering or transfer.  Double-folding
potentials are often used for the real part of the optical
potential. In this first study, we follow the S\~ao Paulo group and
assume the imaginary part of the optical potential $U_{\rm F}$ to be
proportional to its real part~\cite{Pere09ImDFPot}
\begin{equation}
U_{\rm F}(r,E_\text{cm}) = (1+ i \, N_W) \, V_{\rm F}(r,E_\text{cm}) \,, \label{eq:U}
\end{equation}
where $V_{\rm F}$ is our double-folding potential and $N_W$ is a real
coefficient taken in the range 0.6--0.8.

\begin{figure}[t]
\includegraphics[width=\columnwidth]{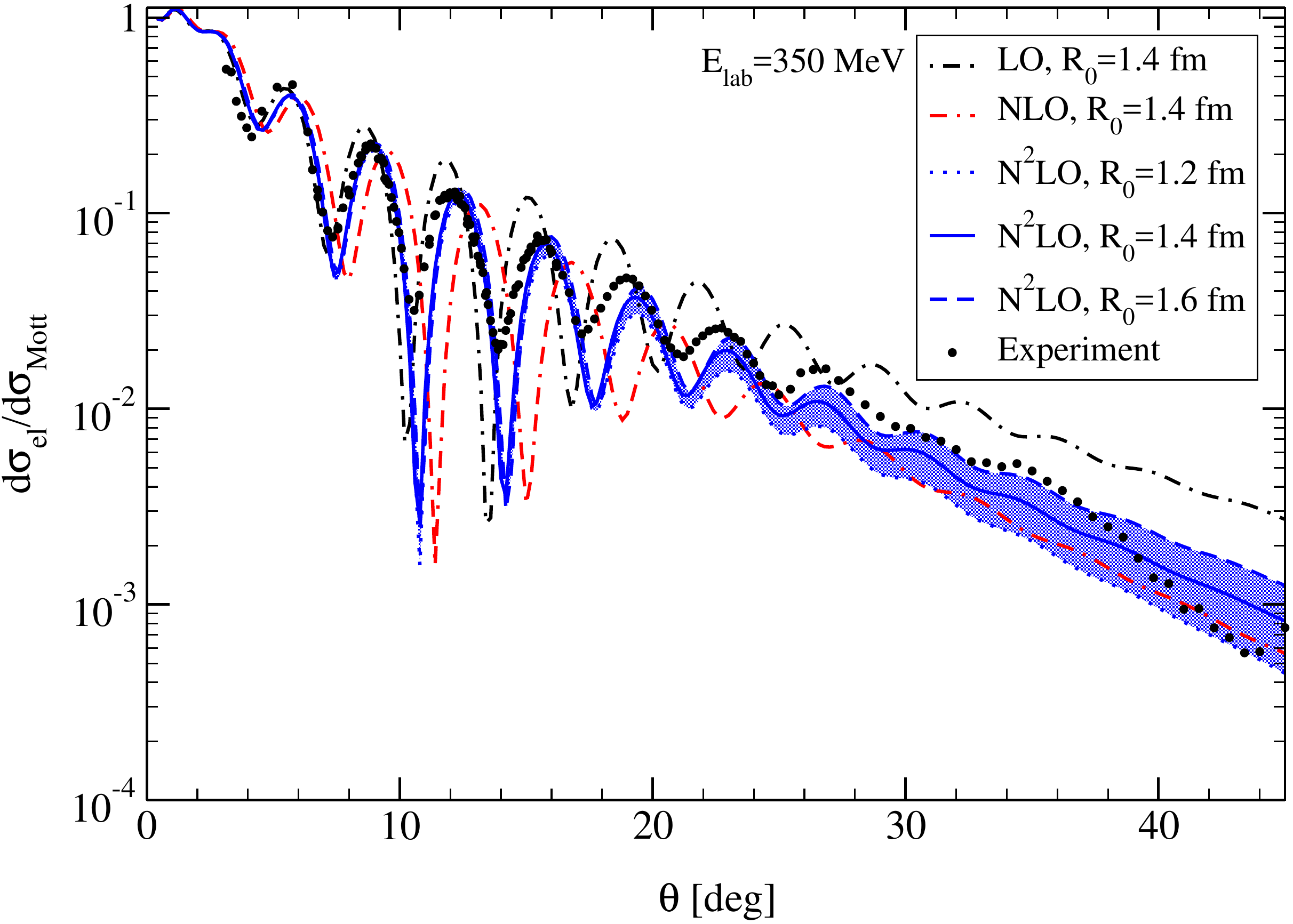}
\caption{Ratio of the cross section for elastic $^{16}$O--$^{16}$O
  scattering to the Mott cross section for laboratory energy
  $E_{\text{lab}}=350$~MeV. Results are shown at LO, NLO, and N$^2$LO
  for $R_0 = 1.4$~fm, and the sensitivity to $R_0 = 1.2$--$1.6$~fm is
  illustrated at N$^2$LO by the shaded area. In all cases, we take for the imaginary part
  $N_W = 0.8$ [see \Eq{eq:U}]. The results are compared to
  experimental data from Ref.~\cite{Bohl93expel}.}
\label{fig:elastic_350}
\end{figure}

The cross section for $^{16}$O--$^{16}$O elastic scattering for
laboratory energy $E_{\rm lab}=350$~MeV is shown in
\Fig{fig:elastic_350} as a ratio to the Mott cross section.  In these
calculations, we take for the imaginary part $N_W = 0.8$, whereas we
study the sensitivity to $N_W$ later. Note that since $^{16}$O is a
spinless boson, the wave function for the $^{16}$O--$^{16}$O relative
motion needs to be properly symmetrized.

\begin{figure}[t]
\includegraphics[width=\columnwidth]{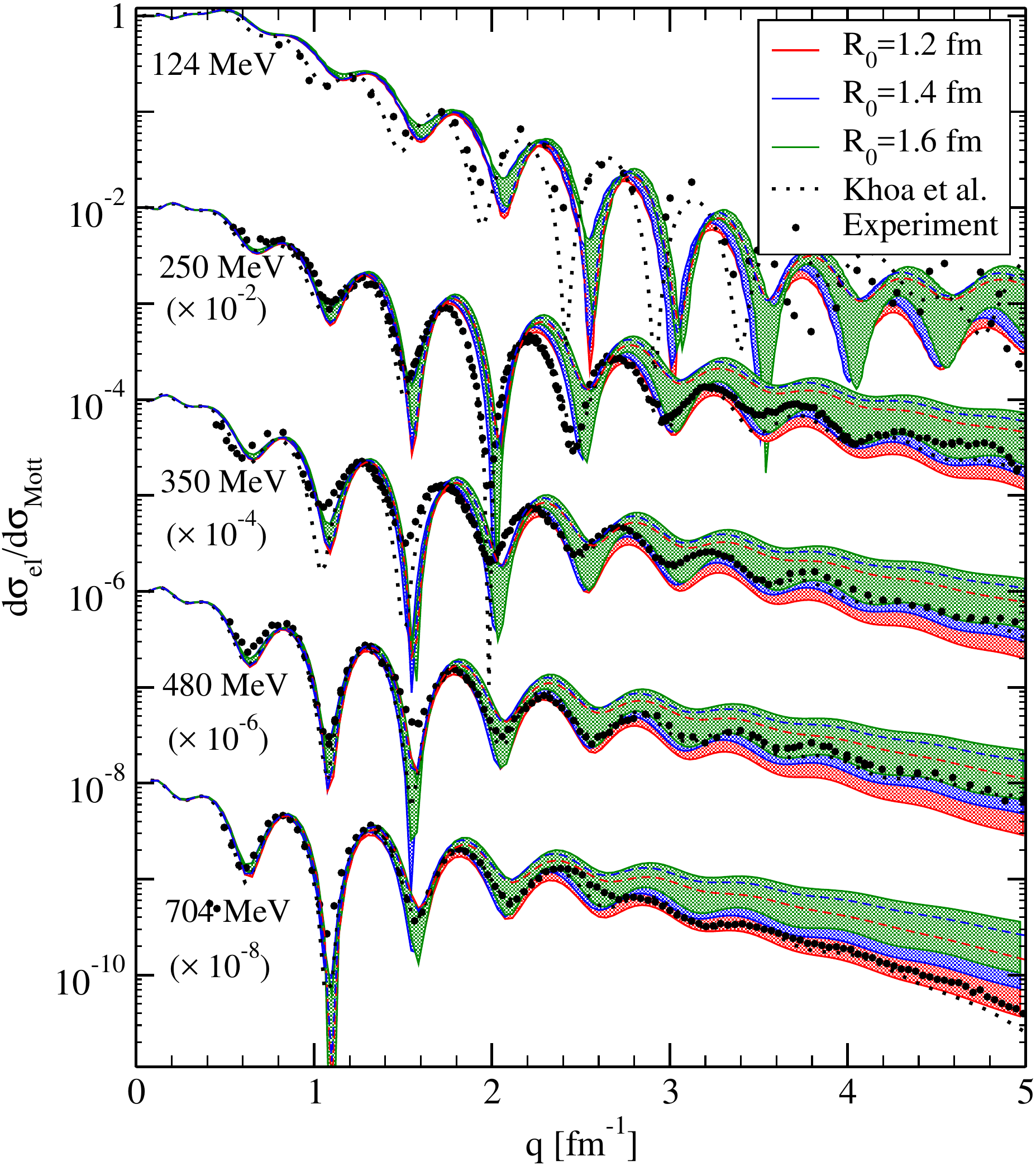}
\caption{Ratio of the cross section for elastic $^{16}$O--$^{16}$O
  scattering to the Mott cross section as a function of momentum
  transfer~$q$ for various laboratory energies (the different energy
  results are offset by a factor as indicated).  Results are shown at
  N$^2$LO for $R_0 = 1.2$~fm (red), 1.4~fm (blue), and 1.6~fm
  (green). For these cutoffs, the region between the results with
  $N_W=0.6$ (upper limit) and $N_W=0.8$ (lower limit) is shaded. In
  the case of $R_0 = 1.2$~fm and 1.4~fm, the upper line is shown as a
  dashed line.  For comparison, we also show the optical-potential
  results of Khoa {\it et al.}~\cite{Khoa0016OEl} and the experimental
  data from
  Refs.~\cite{Bohl93expel,Kond96expel,Bart96expel,Nuof98expel,Khoa0016OEl,Nico99expel}.}
\label{fig:elastic_q}
\end{figure}

As in Figs.~\ref{fig:Phase_shifts} and \ref{fig:VF_350}, we observe a
systematic order-by-order behavior. The uncertainty related to the
cutoff choice at N$^2$LO (shaded area) is similar to that observed in
the double-folding potential itself (see \Fig{fig:VF_350}). At forward
angles, i.e., up to $10^\circ$, the agreement of our calculations with
experiment is excellent, knowing in particular that there are no
parameters fitted to reproduce the data. At larger angles this
agreement deteriorates. Since the spread observed in the NN cutoff
band remains small even at larger angles, this discrepancy cannot be
fully explained by the detail of the $NN$ interactions considered.  It
is likely due to the simple Hartree-Fock level of the many-body
calculation or to the choice of the $^{16}$O density, which could be
improved. In addition, it could also reflect the simple description of
the imaginary part.  The elastic scattering cross sections computed at
various laboratory energies between 124 and 704~MeV are displayed in
\Fig{fig:elastic_q} as a ratio to the Mott cross section.  To compare
the calculations performed at different energies, we plot them as a
function of the momentum transfer~$q$. The bands are delimited by
results for the range $N_W=0.6-0.8$. Results generated by the cutoffs
$R_0=1.2$~fm, 1.4~fm, and 1.6~fm are displayed in red, blue, and
green, respectively. We find that the cutoff variation is less
relevant than the impact of the imaginary part coefficient $N_W$.  As
in \Fig{fig:elastic_350}, we observe a general agreement between our
calculations and the data, especially at forward angles. At larger
momentum transfer, the agreement is less good, although the
experimental points remain close to the spread obtained for the $N_W$
range.  This confirms that going beyond the simple description of the
imaginary part could improve our calculations.

For comparison, we also show the cross sections computed with the
phenomenological optical potential developed by Khoa {\it et
  al.}~\cite{Khoa0016OEl} (dotted line in \Fig{fig:elastic_q}).  This
potential, containing nine adjustable parameters that are modified at
each energy, provides a near-perfect reproduction of the data. Given
that we do not include any adjustable parameter to fit the data, our
results with the double-folding potential based on chiral EFT
interactions are therefore very encouraging.

\section{Fusion reactions}
\label{sec:S_fac}

The $^{16}$O+$^{16}$O fusion reaction is another test for our
double-folding potential.  This cross section $\sigma_{\text{fus}}$
has been measured at low energies to study the role of intermediate
resonances during fusion~\cite{Tser78expfus,Hulk80expfus} and because
this reaction takes place in medium- to heavy-mass
stars~\cite{Hulk80expfus,Kuro87expfus,Wu84expfus,Duar15expfus}.
Oxygen fusion is crucial in medium-mass nuclei burning chains, which
provide the seeds to the synthesis of heavy elements. At low energy,
the reaction takes place through quantum tunneling of the effective
potential barrier that results from the combination of the attractive
strong interaction, the repulsive Coulomb interaction, and the
centrifugal term of the kinetic energy:
\begin{equation}
V_\text{eff}(r,E_\text{cm}) = V_\text{F}(r,E_\text{cm}) + V_\text{Coul}(r) + \frac{l(l+1)}{2\mu r^2} \,.
\end{equation}
Since the fusion reaction takes place at very low energies and
involves light spherical nuclei, we take the (real) double-folding
potential as the nuclear interaction for this
reaction~\cite{Hagi12IWBC}. For light systems like $^{16}$O+$^{16}$O,
the fusion barrier is at around 9~fm, well before the neck formation,
which justifies the use of the double-folding procedure.  For the code
used in the computation of the fusion cross section, we approximate
the Coulomb interaction by a sphere-sphere potential of radius $R_C =
2 \times 4.39$~fm~\cite{Baye82SphCoul}. We do not expect this change
from the double-folding Coulomb term used in \Eq{eq:k} to affect
significantly our results.

The fusion cross section of $^{16}$O+$^{16}$O can be obtained from the
probability $P_l$ to tunnel through the barrier in each of the partial
waves~\cite{Hagi12IWBC}
\begin{equation}
\sigma_{\text{fus}}(E_\text{cm})=\frac{\pi}{k^2}\sum_{l}(1+(-1)^l)(2l+1)P_l(E_\text{cm})\,.
\end{equation}
The probabilities $P_l$ are determined using the incoming-wave
boundary condition detailed in Ref.~\cite{Hagi12IWBC} and implemented
in the code CCFULL~\cite{Hagi99IWBCcode}, in which we have included
the effects of the symmetrization of the wave function for the fusing
nuclei being identical spinless bosons.

At low energy, the fusion process is strongly hindered by the Coulomb
repulsion.  This effect is well accounted for by the Gamow factor,
which is usually factorized out of the cross section to define the
astrophysical $S$ factor
\begin{equation}
S(E_\text{cm})=E_\text{cm} \, e^{2\pi\eta} \, \sigma_{\text{fus}}(E_\text{cm}) \,,
\end{equation}
where the Sommerfeld parameter is given by $\eta = Z_1 Z_2 e^2/(4 \pi
\varepsilon_0 v)$, with $v$ the relative velocity between the two
nuclei.

\begin{figure}[t]
\includegraphics[width=\columnwidth]{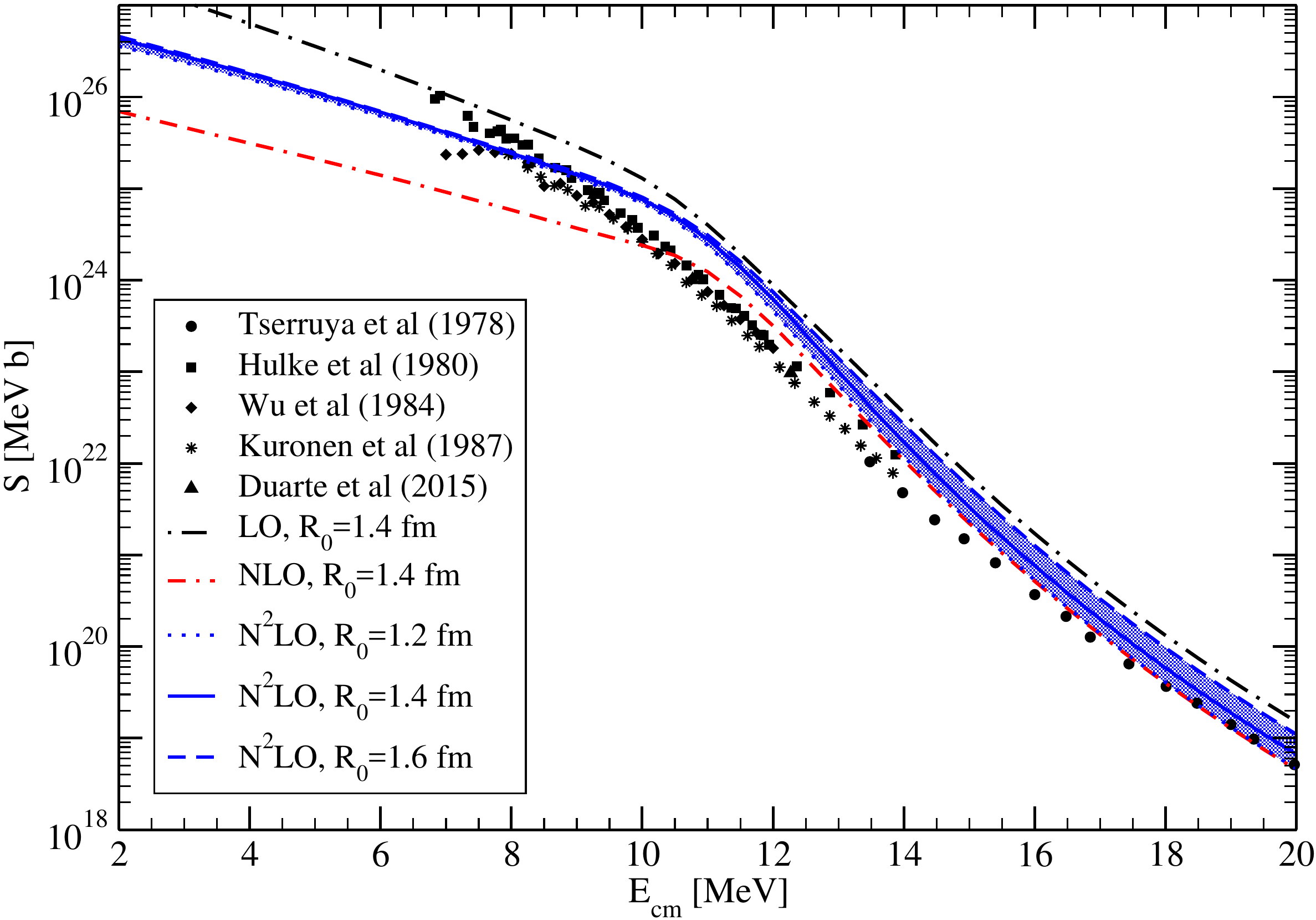}
\caption{Astrophysical $S$-factor for the fusion of $^{16}$O+$^{16}$O
  as a function of the energy $E_{\rm cm}$ in the center-of-mass
  system. Results are shown at LO, NLO, and N$^2$LO for $R_0 =
  1.4$~fm, and the sensitivity to $R_0 = 1.2$--$1.6$~fm  
  at N$^2$LO is illustrated by the shaded area. The results are compared to experimental data from
  Refs.~\cite{Tser78expfus,Hulk80expfus,Wu84expfus,Kuro87expfus,Duar15expfus}.}
\label{fig:S_factor}
\end{figure}

The $S$ factor obtained at LO, NLO, and N$^2$LO for $R_0 = 1.4$~fm and
with different cutoffs $R_0$ at N$^2$LO is displayed in
Fig.~\ref{fig:S_factor}.  Given the very weak energy dependence of the
double-folding potential observed at the relevant energies,
$V_{\text{Ex}}$ is taken at the center of the energy range,
$E_\text{cm} = 12$~MeV. We have tested that taking a different energy
in this range leads to indistinguishable results from those in
Fig.~\ref{fig:S_factor}.  It is interesting to note that, due to the nearly cutoff-independent asymptotic behavior of the nuclear folding potential, the spread between the results obtained with different values of $R_0$ is small around the Coulomb barrier. This leads to results at N$^2$LO in
\Fig{fig:S_factor} that are closer than what \Fig{fig:VF_350} would
suggest. Note also that the less attractive potentials (at
NLO with $R_0=1.4$~fm, and N$^2$LO with $R_0=1.2$~fm) naturally lead
to the lowest fusion cross sections. The general agreement with the
data is good, recalling that there is no fitting parameter. As for
elastic scattering, we observe that the sensitivity to the details in
the $NN$ interaction shown by the shaded area can only partially
explain the discrepancy with experiment. In future work, we will
explore how a better many-body calculation of the double-folding
potential and more realistic densities of the fusing nuclei may
improve this agreement.

\section{Summary and outlook}
\label{sec:sum}

We have presented a first study of constructing nucleus-nucleus
potentials from local chiral $NN$
interactions~\cite{Geze13QMCchi,Geze14long,Huth17Fierz} using the
double-folding method applied to the $^{16}$O--$^{16}$O system. Our
results show that for soft cutoffs, $R_0 \gtrsim 1.4$~fm, the
resulting double-folding potential exhibits a systematic
order-by-order behavior expected in EFT and a weak cutoff dependence
on the details of the $NN$ interactions used. These features carry
through to the elastic scattering cross section and the $S$-factor for
the fusion reaction.

We have focused on the $^{16}$O--$^{16}$O reactions, because these
have been accurately measured and are well studied
theoretically~\cite{Stil89expel,Bohl93expel,Sugi93expel,Kond96expel,Bart96expel,Nuof98expel,Nico99expel,Khoa0016OEl,Tser78expfus,Hulk80expfus,Wu84expfus,Kuro87expfus,Duar15expfus,Kond9016ODeepPot}.
In all cases, a good agreement with the data has been obtained without
any fitting parameter.  Our results thus suggest that the idea to
derive nucleus-nucleus potentials using the double-folding method
based on local chiral EFT interactions is very promising.

We consider this a first step in a more fundamental description of
nucleus-nucleus potentials, but there are several directions how the
calculations can be improved, both at the level of the input
interactions and the many-body folding method. First, the influence of
the nucleon density of the colliding nuclei needs to be evaluated.
This can be done by using more realistic densities, such as those
obtained from electron-scattering measurements or accurate
nuclear-structure models. Second, we need to refine the imaginary part
of the potential.  Assuming it to be proportional to the
double-folding potential provides a first estimate, but it is clear
that this can be improved. Comparisons with phenomenological
potentials~\cite{Khoa0016OEl} and potentials built from inversion
techniques~\cite{Alle93OpPotInv16O,Benn96InvSctdat} can also provide
tests towards more realistic prescriptions.  In a calculation beyond
Hartree-Fock, an imaginary part as well as nonlocal contributions
would arise (see, e.g., Refs.~\cite{Fesh58,Fesh62}).  Moreover, going
beyond the level of the density-matrix expansion considered here,
there will be gradient corrections~\cite{Nege72DME1} (i.e., surface
terms) to the double-folding potential. Finally three-nucleon
interactions need to be investigated in this approach, as they also
enter at N$^2$LO.

In conclusion, coupling chiral EFT interactions with the
double-folding method provides nucleus-nucleus potentials that lead to
very encouraging agreement with elastic-scattering and fusion data in
a broad range of energies. This idea is thus a promising first step
towards the construction of microscopic optical potentials from first
principles with control over uncertainty estimates. Through the above
future developments, we hope to improve this new method to obtain a
systematic way to build efficient optical potentials for nuclear
reactions.

\section*{Acknowledgments}

We thank J.\ E.\ Lynn and I.\ Tews for many helpful discussions, and in
particular for providing the code for the local chiral interactions as
well as for discussions on the softer local chiral interactions
presented here. We also thank D.\ Baye for helpful discussions on the
reactions studied in this work, and the International Atomic Energy
Agency that provided the experimental data through their web page
\href{www-nds.iaea.org}{www-nds.iaea.org}. This work was supported in
part by the European Research Council Grant No. 307986 STRONGINT, the
GSI-TU Darmstadt cooperation, the European Union's Horizon 2020
research and innovation programme under Grant Agreement No. 654002,
and the US National Science Foundation Grant No. PHY-1514695.

\bibliography{references}

\end{document}